\documentclass[natbib,authoryear,preprint]{elsarticle}
\usepackage{graphicx}

\begin{document}

\begin{frontmatter}

\title{Realistic Thermodynamic and Statistical-Mechanical Measures for Neural Synchronization}

\author{Sang-Yoon Kim}
\ead{sykim@labasis.com}
\address{Research Division, LABASIS Corporation, Chunchon, Gangwon-Do 200-702, Korea}
\author{Woochang Lim\corref{wclim}}
\ead{woochanglim@dnue.ac.kr}
\cortext[wclim]{Corresponding Author.}
\address{Department of Science Education, Daegu National University of Education, Daegu 705-115, Korea}

\begin{abstract}
Synchronized brain rhythms, associated with diverse cognitive functions, have been observed in electrical recordings of brain activity. Neural synchronization may be well described by using the population-averaged global potential $V_G$ in computational neuroscience. The time-averaged fluctuation of $V_G$ plays the role of a ``thermodynamic'' order parameter $\cal {O}$ used for describing the synchrony-asynchrony transition in neural systems. Population spike synchronization may be well visualized in the raster plot of neural spikes. The degree of neural synchronization seen in the raster plot is well measured in terms of a ``statistical-mechanical'' spike-based measure $M_s$ introduced by considering the occupation and the pacing patterns of spikes. The global potential $V_G$ is also used to give a reference global cycle for the calculation of $M_s$. Hence, $V_G$ becomes an important collective quantity because it is associated with calculation of both $\cal {O}$  and $M_s$. However, it is practically difficult to directly get $V_G$ in real experiments. To overcome this difficulty, instead of $V_G$, we employ the instantaneous population spike rate (IPSR) which can be obtained in experiments, and develop realistic thermodynamic and statistical-mechanical measures, based on IPSR, to make practical characterization of the neural synchronization in both computational and experimental neuroscience. Particularly, more accurate characterization of weak sparse spike synchronization can be achieved in terms of realistic statistical-mechanical IPSR-based measure, in comparison with the conventional measure based on $V_G$.
\end{abstract}


\end{frontmatter}

\section{Introduction}
\label{sec:INT}
Recently, much attention has been paid to brain rhythms observed in scalp electroencephalogram and local field potentials \citep{Buz}. These brain rhythms emerge via synchronization between individual firings in neural circuits. This kind of neural synchronization may be used for efficient sensory and cognitive processing such as sensory perception, multisensory integration, selective attention, and memory formation \citep{Wang1,Wang2,Gray}, and it is also correlated with pathological rhythms associated with neural diseases (e.g., epileptic seizures and tremors in the Parkinson's disease) \citep{TW}. Here, we are interested in characterization of these synchronized brain rhythms \citep{M1,M2,M3}.

A neural circuit in the major parts of the brain is composed of a few types of excitatory principal cells and diverse types of inhibitory interneurons. By providing a coherent oscillatory output to the principal cells, interneuronal networks play the role of the backbones of many brain rhythms \citep{Buz,Wang1,Wang2,Inhibitory}. In this paper, we consider an inhibitory population of fast spiking (FS) Izhikevich subthreshold interneurons \citep{Izhi1,Izhi2,Izhi3,Izhi4}. Sparsely synchronized neural oscillations are found to appear in an intermediate range of noise intensity. At the population level, fast synchronized rhythms emerge, while at the cellular level, individual neurons discharge stochastic firings at low rates than the population frequency. Fast cortical rhythms [e.g., beta (15-30 Hz), gamma (30-100 Hz), and ultrafast (100-200 Hz) rhythms], associated with diverse cognitive functions, typically exhibit sparse synchronization \citep{Wang1,Sparse1,Sparse2,Sparse3,Sparse4,Sparse5,Sparse6}. The main purpose of our work is to make practical characterization of synchronized cortical rhythms by using realistic measures applicable in both computational and experimental neuroscience.

Neural synchronization may be well described in terms of the population-averaged global potential $V_G$ in computational neuroscience. For a synchronous case, an oscillating global potential $V_G$ appears; otherwise (i.e., $V_G$ is stationary) the population state becomes unsynchronized. Thus, the mean square deviation of $V_G$ plays the role of an order parameter $\cal {O}$ used for describing the synchrony-asynchrony transition in neural systems \citep{Order1,Order2,Order3,Order4,Order5,Kim1,Kim2,Kim3,Kim4}. The order parameter $\cal {O}$ can be regarded as a ``thermodynamic'' measure because it concerns just the the macroscopic global potential $V_G$ without considering any quantitative relation between $V_G$ and the microscopic individual potentials. Through calculation of $\cal {O}$, one can determine the region of noise intensity where synchronized rhythms appear. Population spike synchronization may be well visualized in the raster plot of neural spikes (i.e., a spatiotemporal plot of neural spikes) which can be directly obtained in experiments. For the synchronous case, ``stripes" (composed of spikes and indicating population synchronization) are found to be formed in the raster plot. Due to synchronous contribution of spikes, local maxima of the global potential $V_G$ appear at the centers of stripes. Recently, a ``statistical-mechanical'' spike-based measure $M_s$ was introduced by taking into consideration both the occupation pattern and the pacing pattern of spikes in the stripes of the raster plot \citep{Kim1,Kim2,Kim3,Kim4}. Particularly, the pacing degree between spikes is determined in a statistical-mechanical way by quantifying the average contribution of (microscopic) individual spikes to the (macroscopic) global potential $V_G$. The global potential $V_G$ is thus used to provide a reference global cycle for the calculation of both the occupation and the pacing degrees. Hence, $V_G$ becomes an important population-averaged quantity because it is involved in calculation of both $\cal {O}$ and $M_s$. However, to directly obtain $V_G$ in real experiments is very difficult. To overcome this difficulty, instead of $V_G$, we use an experimentally-obtainable instantaneous population spike rate (IPSR) which is often used as a collective quantity showing population behaviors \citep{Wang1,Sparse1,Sparse2,Sparse3,Sparse4,Sparse5,Sparse6}, and develop realistic thermodynamic and statistical-mechanical measures, based on IPSR, to make practical characterization of the neural synchronization in both computational and experimental neuroscience.
These realistic thermodynamic and statistical-mechanical measures are in contrast to conventional ``microscopic'' synchronization measures such as the correlation-based measure (based on the cross-correlation between the microscopic individual potentials of pairs of neurons) \citep{KI2,CM} and the spike-based measures (based on the spike-distance \citep{SD1,SD2,SD3,SD4,SD5} and the ISI(interspike interval)-distance \citep{ISID} between the microscopic individual spike trains of pairs of neurons). The correlation-based and the spike-based measures are microscopic ones because both of them concern just the microscopic individual potentials or spike-trains without taking into account any quantitative relation between the microscopic quantities and the global activities (e.g., $V_G$ and IPSR). In addition to characterization of population spike synchronization, the conventional spike-based measures are also used to quantify the reliability of spike timing \citep{ST1,ST2,ST3,ST4,ST5,ST6,ST7,ST8,ST9} and the reliability of stimulus discrimination \citep{STD1,STD2,Song1,Song2,STD3,SD}.

This paper is organized as follows. In Sec.~\ref{sec:Izhikevich}, we describe a biological globally-coupled network composed of FS Izhikevich subthreshold neurons. The Izhikevich neurons are not only biologically plausible, but also computationally efficient \citep{Izhi1,Izhi2,Izhi3,Izhi4}, and they interact through inhibitory GABAergic synapses (involving the $\rm {GABA_A}$ receptors). In Sec.~\ref{sec:Measure}, we develop realistic thermodynamic and statistical-mechanical measures, based on IPSR, which are applicable in both the computational and the experimental neuroscience. Their usefulness for characterization of neural synchronization is shown in explicit examples. Through calculation of the realistic thermodynamic order parameter, we determine the range of noise intensity where sparsely synchronized neural oscillations occur. In the synchronous region of noise intensity, we also characterize synchronized rhythms in terms of realistic statistical-mechanical spiking measure $M_s$. It is thus shown that $M_s$ is effectively used to characterize sparse synchronization shown in partially-occupied stripes of the raster plot. 
For examination on effectiveness of realistic statistical-mechanical measure $M_s$, we have also successfully characterized neural synchronization in another population of FS Wang-Buzs\'{a}ki suprathrshold interneurons \citep{KI2} in Section \ref{sec:WB}. 
Furthermore, it has been shown that more accurate characterization of weak sparse spike synchronization can be achieved in terms of statistical-mechanical IPSR-based measures, in comparison with the conventional statistical-mechanical $V_G$-based measures.
Finally, a summary along with discussion on the applicability of realistic statistical-mechanical measure to real experimental data is given in Section \ref{sec:SUM}.

\section{Inhibitory Population of FS Izhikevich Subthreshold Neurons}
\label{sec:Izhikevich}

We consider an inhibitory population of $N$ globally-coupled subthreshold neurons. As an element in our coupled neural system, we choose the FS Izhikevich interneuron model which is not only biologically plausible, but also computationally efficient \citep{Izhi1,Izhi2,Izhi3,Izhi4}. The population dynamics in this neural network is governed by the following set of ordinary differential equations:
\begin{eqnarray}
C\frac{dv_i}{dt} &=& k (v_i - v_r) (v_i - v_t) - u_i +I_{DC} +D \xi_{i} -I_{syn,i}, \label{eq:CIZA} \\
\frac{du_i}{dt} &=& a \{ U(v_i) - u_i \}, \label{eq:CIZB} \\
\frac{ds_i}{dt}&=& \alpha s_{\infty}(v_i) (1-s_i) - \beta s_i, \;\;\; i=1, \cdots, N, \label{eq:CIZC}
\end{eqnarray}
with the auxiliary after-spike resetting:
\begin{equation}
{\rm if~} v_i \geq v_p,~ {\rm then~} v_i \leftarrow c~ {\rm and~} u_i \leftarrow u_i + d, \label{eq:RS}
\end{equation}
where
\begin{eqnarray}
U(v) &=& \left\{ \begin{array}{l} 0 {\rm ~for~} v<v_b \\ b(v - v_b)^3 {\rm ~for~} v \ge v_b \end{array} \right. , \label{eq:CIZD} \\
I_{syn,i} &=& \frac{J}{N-1} \sum_{j(\ne i)}^N s_j(t) (v_i - V_{syn}), \label{eq:CIZE} \\
s_{\infty} (v_i) &=& 1/[1+e^{-(v_i-v^*)/\delta}]. \label{eq:CIZF}
\end{eqnarray}
Here, the state of the $i$th neuron at a time $t$ is characterized by three state variables: the membrane potential $v_i$, the recovery current $u_i$ and the synaptic gate variable $s_i$ denoting the fraction of open synaptic ion channels. In Eq.~(\ref{eq:CIZA}), $C$ is the membrane capacitance, $v_r$ is the resting membrane potential, and $v_t$ is the instantaneous threshold potential. After the potential reaches its apex (i.e., spike cutoff value) $v_p$, the membrane potential and the recovery variable are reset according to Eq.~(\ref{eq:RS}). The units of the capacitance $C$, the potential $v$, the current $u$ and the time $t$ are pF, mV, pA, and ms, respectively.

Unlike Hodgkin-Huxley-type conductance-based models, the Izhikevich model matches neuronal dynamics by tuning the parameters instead of matching neuronal electrophysiology. The parameters $k$ and $b$ are associated with the neuron's rheobase and input resistance, $a$ is the recovery time constant, $c$ is the after-spike reset value of $v$, and $d$ is the total amount of outward minus inward currents during the spike and affecting the after-spike behavior (i.e., after-spike jump value of $u$). Tuning these parameters, the Izhikevich neuron model may produce 20 of the most prominent neuro-computational features of cortical neurons \citep{Izhi1,Izhi2,Izhi3,Izhi4}. Here, we consider the FS Izhikevich interneurons. These FS interneurons do not fire postinhibitory (rebound) spikes, and hence they are simulated with nonlinear $u$-nullcline of $U(v)=0$ \citep{Izhi3}. Here, we use the parameter values for the FS interneurons in the layer 5 Rat visual cortex \citep{Izhi3};
$C=20,~v_r=-55,~v_t=-40,~v_p=25,~v_b=-55,~k=1,~a=0.2,~b=0.025,~c=-45,~d=0.$	

Each Izhikevich interneuron is stimulated by using the common DC current $I_{DC}$ and an independent Gaussian white noise $\xi_i$ [see the 3rd and the 4th terms in Eq.~(\ref{eq:CIZA})] satisfying $\langle \xi_i(t) \rangle =0$ and $\langle \xi_i(t)~\xi_j(t') \rangle = \delta_{ij}~\delta(t-t')$, where
$\langle\cdots\rangle$ denotes the ensemble average. The noise $\xi$ is a parametric one that randomly perturbs the strength of the applied current
$I_{DC}$, and its intensity is controlled by using the parameter $D$. The FS Izhikevich interneurons exhibit the type-II excitability \citep{Izhi3}. As $I_{DC}$ passes a threshold in the absence of noise, each single type-II Izhikevich interneuron begins to fire with a nonzero frequency that is relatively insensitive to the change in $I_{DC}$ \citep{Ex1,Ex2}. Here, we consider the subthreshold case of $I_{DC}=72$ pA where single neurons cannot fire spontaneously without noise.

The last term in Eq.~(\ref{eq:CIZA}) represents the synaptic coupling of the network. $I_{syn,i}$ of Eq.~(\ref{eq:CIZE}) represents a synaptic current injected into the $i$th neuron. Here the coupling strength is controlled by the parameter $J$ and $V_{syn}$ is the synaptic reversal potential.                 Here, we use $V_{syn}=-80$ mV for the inhibitory synapse. The synaptic gate variable $s$ obeys the 1st order kinetics of Eq.~(\ref{eq:CIZC}) \citep{Order3,KI2}. Here, the normalized concentration of synaptic transmitters, activating the synapse, is assumed to be an instantaneous sigmoidal function of the membrane potential with a threshold $v^*$ in Eq.~(\ref{eq:CIZF}), where we set $v^*=0$ mV and $\delta=2$ mV. The transmitter release occurs only when the neuron emits a spike (i.e., its potential $v$ is larger than $v^*$). For the inhibitory GABAergic synapse (involving the $\rm{GABA_A}$ receptors), the synaptic channel opening rate, corresponding to the inverse of the synaptic rise time $\tau_r$, is $\alpha=10$ ${\rm ms}^{-1}$, and the synaptic closing rate $\beta$, which is the inverse of the synaptic decay time $\tau_d$, is $\beta=0.1$ ${\rm ms}^{-1}$ \citep{GABA1,GABA2}. Hence, $I_{syn}$ rises fast and decays slowly.

Numerical integration of Eqs.~(\ref{eq:CIZA})-(\ref{eq:CIZC}) is done using the Heun method \citep{SDE} (with the time step $\Delta t=0.01$ ms), and data for $(v_i,u_i,s_i)$ $(i=1,\dots,N)$ are obtained with the sampling time interval $\Delta t=0.01$ ms. For each realization of the stochastic process, we choose a random initial point $[v_i(0),u_i(0),s_i(0)]$ for the $i$th $(i=1,\dots, N)$ neuron with uniform probability in the range of $v_i(0) \in (-50,-45)$, $u_i(0) \in (10,15)$, and $s_i(0) \in (0.0,0.02)$.

\section{Characterization of Neural Synchronization in Terms of Realistic Thermodynamic and Statistical-Mechanical Measures}
\label{sec:Measure}
In this section, we study collective spike synchronization in an inhibitory population of FS Izhikevich subthreshold interneurons. We develop realistic thermodynamic and statistical-mechanical measures, based on IPSR, which are applicable in both computational and experimental neuroscience, and show their usefulness for characterization of neural synchronization in explicit examples.

\begin{figure}
\includegraphics[width=0.9\columnwidth]{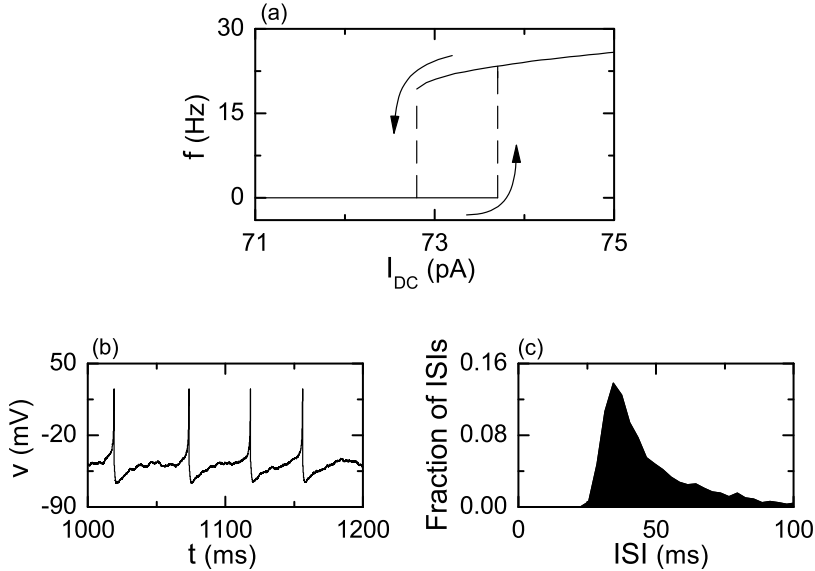}
\caption{Single FS Izhikevich neuron. (a) Plot of the mean firing rate $f$ versus the external DC current $I_{DC}$
for $D=0$. (b) Time series of the membrane potential $v$ and (c) the ISI histogram for $I_{DC}=72$ and $D=20$.
The ISI histogram is made of $5 \times 10^4$ ISIs and the bin size is 3 ms.
}
\label{fig:Single}
\end{figure}

We first consider the case of a single FS Izhikevich interneuron. In the absence of noise (i.e., $D=0$), the Izhikevich interneuron exhibits a jump from a resting state to a spiking state via subcritical Hopf bifurcation for $I_{DC,h}=73.7$ pA by absorbing an unstable limit cycle born via a fold limit cycle bifurcation for $I_{DC,l}=72.8$ pA. Hence, the Izhikevich interneuron shows type-II excitability because it begins to fire with a non-zero frequency that is relatively insensitive to changes in $I_{DC}$, as shown in Fig.~\ref{fig:Single}(a). Throughout this paper, we consider a subthreshold case of $I_{DC}=72$ pA. An isolated subthreshold Izhikevich interneuron cannot fire spontaneously without noise. Figure \ref{fig:Single}(b) shows a time series of the membrane potential $v$ of a subthreshold interneuron for $D=20$ pA $\rm{ms}^{1/2}$. Complex noise-induced spikings appear intermittently. For this subthreshold case, the ISI histogram is shown in Fig.~\ref{fig:Single}(c). The most probable value of the ISIs (corresponding to the main highest peak) is $34.5$ ms (corresponding to 29 Hz). But, due to a long tail in the ISI histogram the average value of the ISIs becomes 47.7 ms, and hence the mean firing rate becomes 21 Hz.

\begin{figure}
\includegraphics[width=0.9\columnwidth]{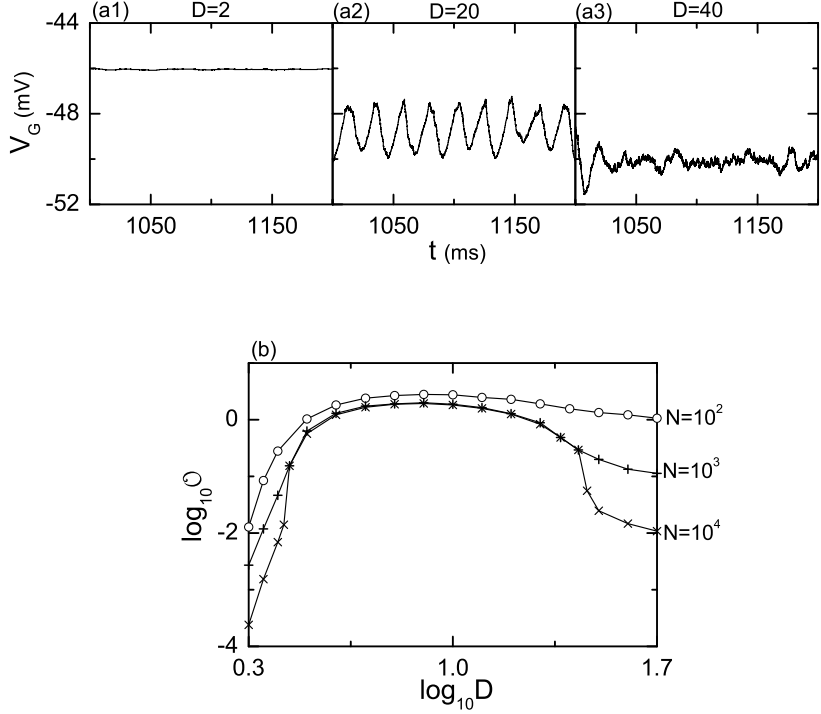}
\caption{Thermodynamic order parameter $\cal{O}$, based on the global potential $V_G$, in an inhibitory population
of $N$ globally-coupled FS Izhikevich subthreshold interneurons for $I_{DC}=72$ and $J=20$. Time series of $V_G(t)$
for $N=10^3$ when (a1) $D=2$, (a2) $D=20$, and (a3) $D=40$. (b) Plots of $\rm{log}_{10}\cal{O}$ versus $\log_{10}D$.
}
\label{fig:Order}
\end{figure}

We consider an inhibitory population of $N$ globally-coupled subthreshold FS Izhikevich interneurons for $I_{DC}=72$ pA and set the coupling strength as $J=20$ nS. (Hereafter, for convenience we omit the dimensions of $I_{DC}$, $J$, and $D$.) By varying the noise intensity $D$, we investigate the population spike synchronization. In computational neuroscience, an ensemble-averaged global potential,
\begin{equation}
 V_G (t) = \frac {1} {N} \sum_{i=1}^{N} v_i(t),
\label{eq:GPOT}
\end{equation}
is often used for describing emergence of population neural synchronization. Figures \ref{fig:Order}(a1)-\ref{fig:Order}(a3) show the time series of $V_G$ for three values of $D$. For a synchronous case, an oscillating global potential $V_G$ appears (e.g., $D=20$), while for un unsynchronized case $V_G$ is nearly stationary (e.g., $D=2$ and 40). Thus, the mean square deviation of the global potential $V_G$,
\begin{equation}
{\cal{O}} \equiv \overline{(V_G(t) - \overline{V_G(t)})^2},
 \label{eq:Order}
\end{equation}
plays the role of an order parameter used for describing the asynchrony-synchrony transition in neural systems \citep{Order1,Order2,Order3,Order4,Order5,Kim1,Kim2,Kim3,Kim4}. (Here the overbar represents the time average.)
This order parameter $\cal{O}$ can be regarded as a thermodynamic measure because it concerns just the macroscopic global potential $V_G$ without taking into consideration any quantitative relation between $V_G$ and the microscopic individual potentials. For the synchronized (unsynchronized) state, the thermodynamic order parameter $\cal{O}$ approaches a non-zero (zero) limit value in the thermodynamic limit of $N \rightarrow \infty$. Figure \ref{fig:Order}(b) shows plots of the order parameter versus the noise intensity. For $D < D^*_l$ $(\simeq 2.6$), unsynchronized states exist because the order parameter $\cal{O}$ tends to zero as $N \rightarrow \infty$. As $D$ passes the lower threshold $D^*_l$, a coherent transition occurs because of a constructive role of noise to stimulate synchronization between noise-induced spikings. However, for large $D > D^*_h$ $(\simeq 29$) such synchronized states disappear (i.e., a transition to an unsynchronized state occurs when $D$ passes the higher threshold $D^*_h$) due to a destructive role of noise to spoil the collective spike synchronization.

\begin{figure}[t]
\includegraphics[width=0.9\columnwidth]{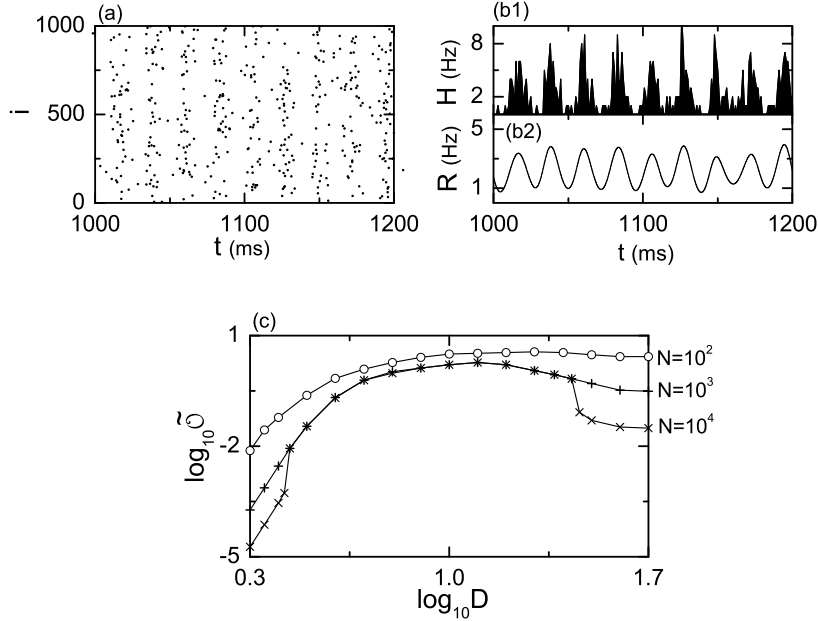}
\caption{Realistic thermodynamic order parameter $\tilde{\cal{O}}$, based on $R(t)$ (IPSR kernel estimate),
in an inhibitory population of $N$ globally-coupled FS Izhikevich subthreshold interneurons for $I_{DC}=72$ and $J=20$.
(a) Raster plot of neural spikes for $N=10^3$ and $D=20$. IPSR (b1) histogram $H(t)$ and (b2) kernel estimate $R(t)$ for $N=10^3$ and $D=20$. The bin size for $H(t)$ is 1 ms and the band width for the Gaussian kernel estimate is 4 ms. (c) Plots of $\rm{log}_{10}\tilde{\cal{O}}$ versus $\log_{10}D$
}
\label{fig:IPSR}
\end{figure}

As shown above, the global potential $V_G$ is an important population-averaged quantity to describe neural synchronization in computational neuroscience. But, it is practically difficult to directly get $V_G$ in real experiments. To overcome this difficulty, instead of $V_G$, we use the IPSR which is an experimentally-obtainable population quantity used in both the experimental and the computational neuroscience \citep{Wang1,Sparse1,Sparse2,Sparse3,Sparse4,Sparse5,Sparse6}. The IPSR is obtained from the raster plot of neural spikes which is a collection of spike trains of individual neurons. Such raster plots of spikes, where population spike synchronization may be well visualized, are fundamental data in experimental neuroscience. For the synchronous case, ``stripes" (composed of spikes and indicating population synchronization) are found to be formed in the raster plot. As an example, Fig.~\ref{fig:IPSR}(a) shows the raster plot of spikes for the case of $D=20$. Nine partially-occupied and smeared stripes, representing sparse population synchronization, are seen in the raster plot. Then the IPSR histogram, $H(t)$, at a time $t$ is given by:
\begin{equation}
H(t) = \frac{N_s(t)}{N \cdot \Delta t},
\label{eq:IPSRH}
\end{equation}
where $\Delta t$ is the bin width for the histogram and $N_s(t)$ is the number of spikes in a bin at time $t$. Figure \ref{fig:IPSR}(b1) shows the IPSR histogram $H(t)$ with bin width $\Delta t =1$ ms. Similar to the global oscillation of $V_G$ [Fig.~2(a2)], $H(t)$ also shows collective oscillatory behavior. But, it seems to be rough. To obtain a smooth IPSR, we employ the kernel density estimation (kernel smoother) \citep{Kernel}.
Each spike in the raster plot is convoluted (or blurred) with a kernel function $K_h(t)$ to obtain a smooth estimate of IPSR, $R(t)$:
\begin{equation}
R(t) = \frac{1}{N} \sum_{i=1}^{N} \sum_{s=1}^{n_i} K_h (t-t_{s}^{(i)}),
\label{eq:IPSRK}
\end{equation}
where $t_{s}^{(i)}$ is the $s$th spiking time of the $i$th neuron, $n_i$ is the total number of spikes for the $i$th neuron, and we use a Gaussian
kernel function of band width $h$:
\begin{equation}
K_h (t) = \frac{1}{\sqrt{2\pi}h} e^{-t^2 / 2h^2}, ~~~~ -\infty < t < \infty.
\label{eq:Gaussian}
\end{equation}
Figure \ref{fig:IPSR}(b2) shows a smooth IPSR kernel estimate $R(t)$ of band width $h=4$ ms. We note that the global oscillation of $R(t)$ seems to be as smooth as that of $V_G(t)$. Hence, instead of $V_G$, we use the IPSR kernel estimate $R(t)$, and develop a realistic thermodynamic order parameter $\tilde{\cal{O}}$, based on $R(t)$:
\begin{equation}
\tilde{\cal{O}} \equiv \overline{(R(t) - \overline{R(t)})^2},
 \label{eq:OrderIPSRK}
\end{equation}
Plots of $\tilde{\cal{O}}$ versus the noise intensity are shown in Fig.~\ref{fig:IPSR}(c). Neural synchronization is found to emerge in a range of
$D^*_l < D < D^*_h$, which is completely consistent with the result obtained through calculation of $\cal{O}$, based on $V_G$ [see Fig.~\ref{fig:Order}(b)]. Thus, $\tilde{\cal{O}}$ becomes a realistic thermodynamic order parameter applicable in both the experimental and the computational neuroscience.

In a synchronous range of $D^*_l < D < D^*_h$, population spike synchronization may be well visualized in the raster plot of neural spikes (spike times of individual Izhikevich neurons correspond to times at which peak potential $v_p$ (=25mV) occurs).
For the synchronous case, ``stripes" (composed of spikes and indicating population synchronization) appear in the raster plot. Recently, a ``statistical-mechanical'' spiking measure $M_s$ was introduced by considering both the occupation and the pacing patterns of spikes in the stripes of the raster plot \citep{Kim1,Kim2,Kim3,Kim4}. The global potential $V_G$ was used to give a reference global cycle for the calculation of both the occupation and the pacing degrees. However, the spiking measure $M_s$, based on $V_G$, is practically inapplicable to the case of experimental neuroscience because to obtain $V_G$ in experiments is difficult. Here, instead of $V_G$, we employ the experimentally-obtainable IPSR kernel estimate $R(t)$, and develop a refined version of statistical-mechanical spiking measure $M_s$, based on $R(t)$, to characterize neural synchronization in both the experimental and the computational neuroscience.

As an example, we consider a synchronous case of $D=20$. As shown in Fig.~\ref{fig:IPSR}(a), the raster plot is composed of partially-occupied and smeared stripes, indicating sparse population synchronization. The spiking measure $M_i$ of the $i$th stripe is defined by the product of the occupation degree $O_i$ of spikes (representing the density of the $i$th stripe) and the pacing degree $P_i$ of spikes (denoting the smearing of the $i$th stripe):
\begin{equation}
M_i = O_i \cdot P_i.
\label{eq:SM}
\end{equation}
The occupation degree $O_i$ of spikes in the stripe is given by the fraction of spiking neurons:
\begin{equation}
   O_i = \frac {N_i^{(s)}} {N},
\end{equation}
where $N_i^{(s)}$ is the number of spiking neurons in the $i$th stripe. For the full occupation $O_i=1$, while for the partial occupation $O_i<1$.
The pacing degree $P_i$ of spikes in the $i$th stripe can be determined in a statistical-mechanical way by taking into account its contribution to the macroscopic IPSR kernel estimate $R(t)$.

\begin{figure}[t]
\includegraphics[width=0.9\columnwidth]{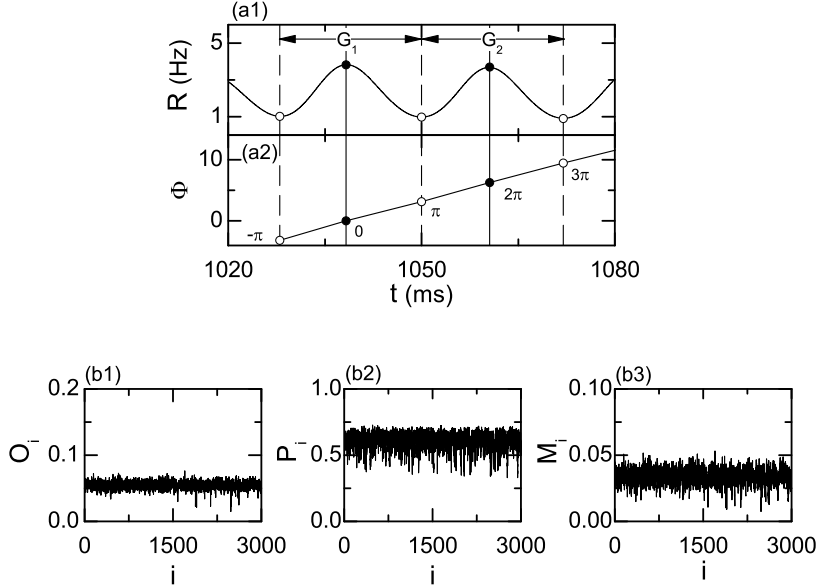}
\caption{Realistic statistical-mechanical spiking measure $M_s$, based on the IPSR kernel estimate $R(t)$, in an inhibitory population of $N$ $(=10^3$) globally-coupled FS Izhikevich subthreshold interneurons for $I_{DC}=72$, $J=20$, and $D=20$. Time series of (a1) the IPSR kernel estimate $R(t)$ and (a2) the global phase $\Phi(t)$. Plots of (b1) $O_i$ (occupation degree of spikes in the $i$th stripe), (b2) $P_i$ (pacing degree of spikes in the $i$th stripe), and (b3) $M_i$(spiking measure in the $i$th stripe) versus $i$ (stripe). In (a1) and (a2), vertical dashed and solid lines represent the times at which local minima and maxima (denoted by open and solid circles) of $R(t)$ occur, respectively and $G_i$ ($i=1,2$) denotes the $i$th global cycle.}
\label{fig:SM1}
\end{figure}

Figure \ref{fig:SM1}(a1) shows a time series of the IPSR kernel estimate $R(t)$; local maxima and minima are denoted by solid and open circles, respectively. Obviously, central maxima of $R(t)$ between neighboring left and right minima of $R(t)$ coincide with centers of stripes in the raster plot. The global cycle starting from the left minimum of $R(t)$ which appears first after the transient time $(=10^3$ ms) is regarded as the 1st one, which is denoted by $G_1$. The 2nd global cycle $G_2$ begins from the next following right minimum of $G_1$, and so on. Then, we introduce an instantaneous global phase $\Phi(t)$ of $R(t)$ via linear interpolation in the two successive subregions forming a global cycle \citep{GP,Kim1}, as shown in Fig.~\ref{fig:SM1}(a2). The global phase $\Phi(t)$ between the left minimum (corresponding to the beginning point of the $i$th global cycle) and the central maximum is given by:
\begin{equation}
\Phi(t) = 2\pi(i-3/2) + \pi \left(
\frac{t-t_i^{(min)}}{t_i^{(max)}-t_i^{(min)}} \right)
 {\rm~~ for~} ~t_i^{(min)} \leq  t < t_i^{(max)}
~~(i=1,2,3,\dots),
\label{eq:PhiL}
\end{equation}
and $\Phi(t)$ between the central maximum and the right minimum (corresponding to the beginning point of the $(i+1)$th cycle) is given by
\begin{equation}
\Phi(t) = 2\pi(i-1) + \pi \left(
\frac{t-t_i^{(max)}}{t_{i+1}^{(min)}-t_i^{(max)}} \right)
 {\rm~~ for~} ~t_i^{(max)} \leq  t < t_{i+1}^{(min)}
~~(i=1,2,3,\dots),
\label{eq:PhiR}
\end{equation}
where $t_i^{(min)}$ is the beginning time of the $i$th global cycle (i.e., the time at which the left minimum of $R(t)$ appears in the $i$th global cycle) and $t_i^{(max)}$ is the time at which the maximum of $R(t)$ appears in the $i$th global cycle. Then, the contribution of the $k$th microscopic spike in the $i$th stripe occurring at the time $t_k^{(s)}$ to $R(t)$ is given by $\cos \Phi_k$, where $\Phi_k$ is the global phase at the $k$th spiking time [i.e., $\Phi_k \equiv \Phi(t_k^{(s)})$]. A microscopic spike makes the most constructive (in-phase) contribution to $R(t)$ when the corresponding global phase $\Phi_k$ is $2 \pi n$ ($n=0,1,2, \dots$) while it makes the most destructive (anti-phase) contribution to $R(t)$ when $\Phi_i$ is $2 \pi (n-1/2)$. By averaging the contributions of all microscopic spikes in the $i$th stripe to $R(t)$, we obtain the pacing degree of spikes in the $i$th stripe:
\begin{equation}
 P_i ={ \frac {1} {S_i}} \sum_{k=1}^{S_i} \cos \Phi_k,
\label{eq:PACING}
\end{equation}
where $S_i$ is the total number of microscopic spikes in the $i$th stripe.
By averaging $M_i$ of Eq.~(\ref{eq:SM}) over a sufficiently large number $N_s$ of stripes, we obtain the realistic statistical-mechanical spiking measure $M_s$, based on the IPSR kernel estimate $R(t)$:
\begin{equation}
M_s =  {\frac {1} {N_s}} \sum_{i=1}^{N_s} M_i.
\label{eq:CM}
\end{equation}
We follow $3 \times 10^3$ stripes and get $O_i$, $P_i$, and $M_i$ in each $i$th stripe, which are are shown in Figs.~\ref{fig:SM1}(b1)-4(b3). Due to sparse discharges of individual neurons, the average occupation degree $\langle O_i \rangle$ $(\simeq 0.054)$, where $\langle \cdots \rangle$ denotes the average over stripes, is very small. Hence, only a fraction (about 1/20) of the total neurons fire in each stripe. On the other hand, the average pacing degree $\langle P_i \rangle$ $(\simeq 0.61)$ is large in contrast to $\langle O_i \rangle$. Consequently, the realistic ``statistical-mechanical'' spiking measure $M_s$ (which represents the population spike synchronization seen in the whole raster plot) is 0.033. The main reason for the low degree of neural synchronization is mainly due to partial occupation. In this way, the realistic statistical-mechanical spiking measure $M_s$ can be used effectively for characterization of sparsely synchronized cortical rhythms because $M_s$ concerns not only the pacing degree, but also the occupation degree of spikes in the stripes of the raster plot.

\begin{figure}[t]
\includegraphics[width=\columnwidth]{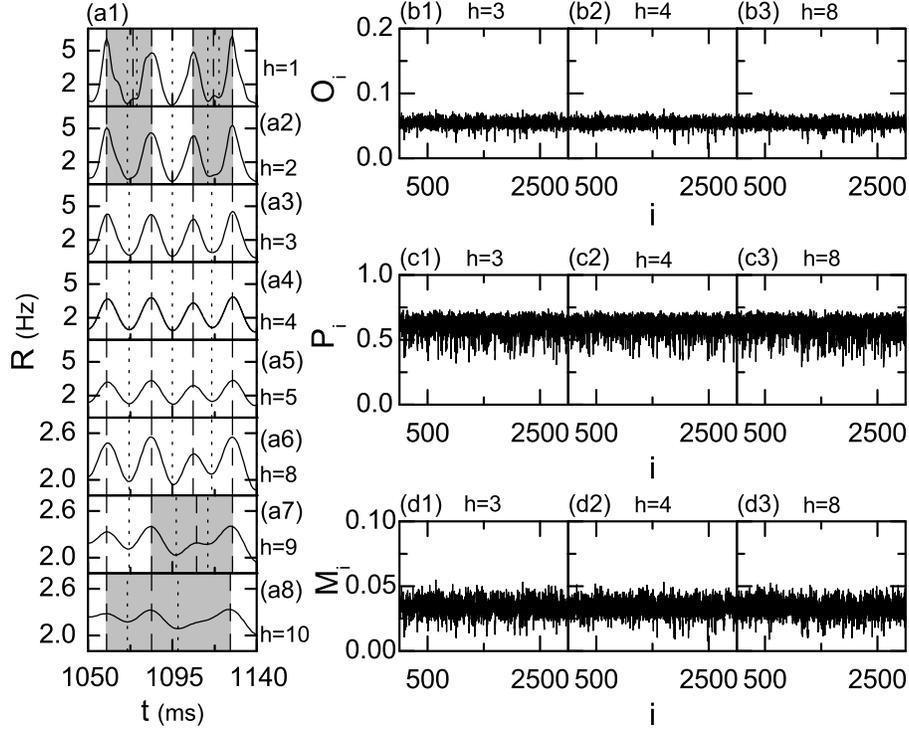}
\caption{Effect of band width $h$ on realistic statistical measures in an inhibitory population of $N$ $(=10^3)$ globally-coupled FS Izhikevich subthreshold interneurons for $I_{DC}=72$, $J=20$ and $D=20$. (a) IPSR kernel estimats for $h=1, 2, 3, 4, 5, 8, 9$, and 10. Vertical dashed and dotted lines represent maxima and minima of the IPSR kernel estimate, respectively. Gray regions in (a1), (a2), (a7), and (a8) represent out-of-phase oscillations of $R(t)$. Plot of $O_i$ (occupation degree of spikes in the $i$th stripe) versus $i$ (stripe) for $h=$ (b1) 3, (b2) 4, and (b3) 8.  Plot of $P_i$ (pacing degree of spikes in the $i$th stripe) versus $i$ (stripe) for $h=$ (c1) 3, (c2) 4, and (c3) 8. Plot of $M_i$ (spiking measure in the $i$th stripe) versus $i$ (stripe) for $h=$ (d1) 3, (d2) 4, and (d3) 8.}
\label{fig:BandWidth}
\end{figure}

The above results in Fig. \ref{fig:SM1} are obtained through IPSR kernel estimate $R(t)$ obtained using Gaussian kernel function of band width $h$=4 ms [see Eqs. (\ref{eq:IPSRK})-(\ref{eq:Gaussian})]. However, these results for $h=4$ ms are still valid in a large range of the band width $h$ ($2<h<9$).
As examples, Figs. \ref{fig:BandWidth}(a1)-\ref{fig:BandWidth}(a8) show $R(t)$ for various values of $h$  for $J=20$ and $D=20$. 
In the valid region of $h$ ($2<h<9$), IPSR kernel estimates $R(t)$ exhibit the ``in-phase" oscillatory behaviors (i.e.,  maxima and minima of $R(t)$ appear synchronously at the same times, independently of $h$), although their amplitudes decrease as $h$ is increased due to band-width effect. On the other hand, outside the valid region
$R(t)$ no longer show in-phase oscillations; see out-of-phase oscillations in gray regions for $h=1, 2, 9$ and 10. 
Consequently, the IPSR kernel estimates in the valid region of $h$ yield the same global phase function $\Phi(t)$ in Eqs. (\ref{eq:PhiL})-(\ref{eq:PhiR}), and the effect of $h$ on realistic statistical-mechanical IPSR-based measures may be neglected, as shown in Figs. \ref{fig:BandWidth}(b1)-\ref{fig:BandWidth}(b3), \ref{fig:BandWidth}(c1)-\ref{fig:BandWidth}(c3), and \ref{fig:BandWidth}(d1)-\ref{fig:BandWidth}(d3). (Hereafter, we continue to fix $h= 4$ ms.)

\begin{figure}[p]
\includegraphics[width=\columnwidth]{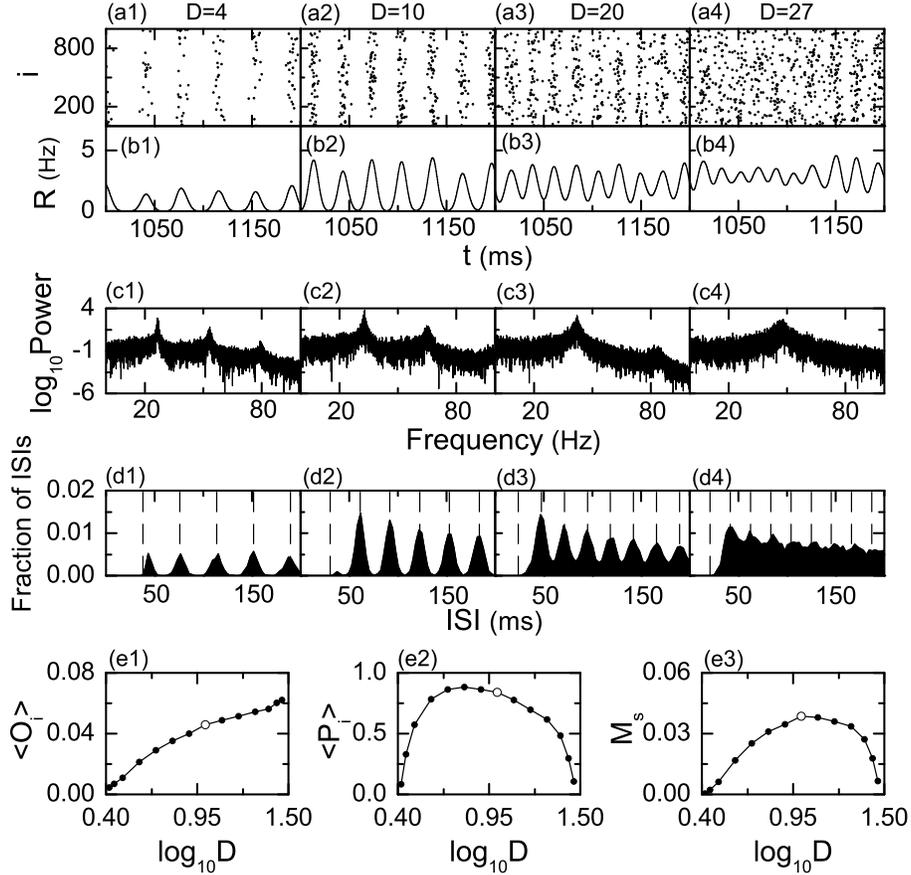}
\caption{Characterization of population spike synchronization in terms of the realistic statistical-mechanical spiking measure $M_s$, based on the IPSR kernel estimate $R(t)$, in an inhibitory population of $N$ $(=10^3)$ globally-coupled FS Izhikevich subthreshold interneurons for $I_{DC}=72$ and $J=20$. (a) Raster plots of neural spikes, (b) IPSR kernel estimates $R(t)$, (c) power spectra, and (d) ISI histograms for (a1)-(d1) $D=4$, (a2)-(d2) $D=10$, (a3)-(d3) $D=20$, and (a4)-(d4) $D=27$. In (c) each power spectrum is made of $2^{16}$ data points. In (d), each ISI histogram is made of $5 \times 10^4$ ISIs and the bin size is 3 ms. Vertical dashed lines in (d) represent the integer multiples of the global period $T_G$ of $R(t)$; $T_G$ = (d1) 37.9 ms, (d2) 30.6 ms, (d3) 23.7 ms, and (d4) 20.8 ms. (e1) Plot of $\langle O_i \rangle$ (average occupation degree of spikes) versus $\log_{10} D$. (e2) Plot of $\langle P_i \rangle$ (average pacing degree of spikes) versus $\log_{10} D$. (e3) Plot of $M_s$ (realistic ``statistical-mechanical'' spiking measure) versus $\log_{10} D$. To obtain $\langle O_i \rangle$, $\langle P_i \rangle$, and $M_s$ in (e1)-(e3), we follow the $3 \times 10^3$ stripes for each $D$. Open circles in (e1)-(e3) denote the data for $D=10$.}
\label{fig:SM2}
\end{figure}

We also vary the noise intensity $D$ in the synchronous region and characterize neural synchronization in terms of the realistic statistical-mechanical spiking measure $M_s$. For the synchronized cases of $D=4$, 10, 20, and 27, Fig.~\ref{fig:SM2} shows the raster plots [(a1)-(a4)], the IPSR kernel estimates $R(t)$ [(b1)-(b4)], the power spectra of $R(t)$ [(c1)-(c4)], and the ISI histograms [(d1)-(d4)] for $N=10^3$.
We measure the degree of population spike synchronization in terms of $\langle O_i \rangle$ (average occupation degree), $\langle P_i \rangle$
(average pacing degree), and $M_s$ (realistic statistical-mechanical IPSR-based spiking measure) for 14 values of $D$ in the synchronized regime, and the results are shown in Figs.~\ref{fig:SM2}(e1)-\ref{fig:SM2}(e3). For the most synchronized case of $D=10$ [where the value of $M_s$ is maximum, as shown in Fig.~\ref{fig:SM2}(e3)], the raster plot in Fig.~\ref{fig:SM2}(a2) is composed of relatively clear partially-occupied stripes with $\langle O_i \rangle$ = 0.046 and $\langle P_i \rangle$ = 0.84  [see open circles in Figs.~\ref{fig:SM2}(e1)-\ref{fig:SM2}(e2)]. This partial occupation occurs due to stochastic spike skipping of individual neurons seen well in the ISI histogram of Fig.~\ref{fig:SM2}(d2) with clear (well-separated) multiple peaks appearing at multiples of the period of $T_G$ (=30.6 ms) of the IPSR kernel estimate $R(t)$. Then, $R(t)$ exhibits relatively regular population oscillation of 32.7 Hz, as shown in Figs.~\ref{fig:SM2}(b2)-\ref{fig:SM2}(c2). As the value of $D$ is increased from 10, the average occupation degree $\langle O_i \rangle$ increases slowly, as might be seen from the raster plots and $R(t)$ in Figs.~\ref{fig:SM2}(a3)-(a4) and \ref{fig:SM2}(b3)-(b4). This slow increase in $\langle O_i \rangle$ is well shown in Fig.~\ref{fig:SM2}(e1). On the other hand, the average pacing degree $\langle P_i \rangle$ for $D>10$ decreases rapidly, as shown in Fig.~\ref{fig:SM2}(e2). For example, stripes in the raster plots of Figs.~\ref{fig:SM2}(a3) and \ref{fig:SM2}(a4) become more and more smeared, and hence the average pacing degree $\langle P_i \rangle$ is decreased with increasing $D$. This smearing of stripes can be understood from the change in the structure of the ISI histograms. As $D$ is increased, peaks begin to merge [see Figs.~\ref{fig:SM2}(d3) and \ref{fig:SM2}(d4)]. This merging of peaks results in the smearing of stripes. Thus, for $D>10$ the degree of population spike synchronization is rapidly decreased as shown in Fig.~\ref{fig:SM2}(e3), mainly due to the rapid decrease in $\langle P_i \rangle$. Consequently, when passing the higher threshold $D^*_h$ $(\simeq 29)$ stripes no longer exist due to complete smearing, and then unsynchronized states appear. In the opposite direction by decreasing the value of $D$ from 10, we also characterize the population spike synchronization. As an example, the raster plot of spikes for $D=4$ is shown in Fig.~\ref{fig:SM2}(a1). The average occupation degree is much decreased to $\langle O_i \rangle=0.022$, as can  be seen in the raster plot. This rapid decrease in $\langle O_i \rangle$ for $D<10$ can be seen in Fig.~\ref{fig:SM2}(e1). On the other hand, the average pacing degree $\langle P_i \rangle$ $(=0.77)$ for $D=4$ is decreased a little when compared with the case of $D=10$; only a little more smearing occurs. However, for $D<4$ both $\langle O_i \rangle$ and $\langle P_i \rangle$ decreases rapidly, as shown in Figs.~\ref{fig:SM2}(e1) and \ref{fig:SM2}(e2). Hence, as $D$ is decreased from 4 the degree of stochastic spiking coherence decreases rapidly. Eventually, when $D$ is decreased through the lower threshold $D^*_l$ $(\simeq 2.6)$, completely scattered sparse spikes appear without forming any stripes in the raster plot, and thus unsynchronized states appear for $D<D^*_l$. In the above way, we characterize neural synchronization in terms of the realistic statistical-mechanical spiking measure $M_s$ in the whole synchronized region, and find that $M_s$ reflects the degree of population spike synchronization seen in the raster plot very well.

\section{Characterization of Neural Synchronization in A Population of Inhibitory Wang-Buzs\'{a}ki Interneurons}
\label{sec:WB}

To examine successful application to an inhibitory population of FS Izhikevich cortical subthreshold interneurons in Section \ref{sec:Measure}, we consider  another population of inhibitory FS Wang-Buzs\'{a}ki suprathreshold interneurons \citep{KI2}:
\begin{eqnarray}
C\frac{dv_i}{dt} &=& -I_{ion,i} +I_{DC} +D \xi_{i} -I_{syn,i}, \label{eq:CWBA} \\
\frac{dx_i}{dt} &=& \phi_x \left [\alpha_x (1-x_i) - \beta_x x_i \right]; \;\;\; x= h \; \textrm{and} \; n, \label{eq:CWBB} \\
\frac{ds_i}{dt} &=& \alpha s_{\infty} (v_i) (1-s_i) - \beta s_i, \;\;\; i=1, \cdots, N, \label{eq:CWBC}
\end{eqnarray}
where
\begin{eqnarray}
I_{ion,i} &=& I_{Na,i} + I_{K,i} + I_{L,i} \label{eq:CWBD} \\
&=& g_{Na} m^{3}_{\infty} h (v_i - V_{Na}) + g_K n^4 (v_i - V_K) +g_L (v_i - V_L), \label{eq:CWBE}  \\
I_{syn,i} &=& \frac{J}{N-1} \sum_{j (\ne i)}^{N} s_j (t) (v_i - V_{syn}), \label{eq:CWBF} \\
\alpha_h &=& 0.07 \cdot e^{-0.05 \cdot (v + 58)}, \label{eq:CWBG} \\
\beta_h &=& 1/[e^{-0.1 \cdot (v+28)}+1], \label{eq:CWBH} \\
\alpha_n &=& [-0.01 \cdot (v+34)] / [e^{-0.1 \cdot (v+34)}-1], \label{eq:CWBI} \\
\beta_n &=& 0.125 \cdot e^{-0.0125 \cdot (v+44)}, \label{eq:CWBJ} \\
m_{\infty} &=& \alpha_m / (\alpha_m + \beta_m), \label{eq:CWBK} \\
\alpha_m &=& [-0.1 \cdot (v+35)] / [e^{-0.1 \cdot (v+35)} - 1], \label{eq:CWBL} \\
\beta_m &=& 4 \cdot e^{-(v+60)/18}, \label{eq:CWBM} \\
s_{\infty} (v_i) &=& 1/[1+e^{-(v_i-v^*)/\delta}]. \label{eq:CWBN}
\end{eqnarray}
Here, the state of the $i$th neuron at a time $t$ (measured in units of ms) is characterized by four state variables: the membrane potential $v_i$ (measured in units of mV), the variable $h_i$ representing the inactivation of the transient $Na^{+}$ current,  the variable $n_i$ representing the activation of the delayed rectifier $K^{+}$ current and the synaptic gate variable $s_i$ denoting the fraction of open synaptic ion channels. In Eq.~(\ref{eq:CWBA}), $C$ is the membrane capacitance, and the time evolution of $v_i$ is governed by four kinds of source currents.
The total ionic current $I_{ion,i}$ of the $i$th neuron consists of the transient sodium current $I_{Na,i}$, the delayed rectifier potassium current $I_{K,i}$ and the leakage current $I_{L,i}$. Each ionic current obeys Ohm’s law. The constants $g_{Na}$, $g_K$ and $g_L$ are the maximum conductances for the ion and the leakage channels, and the constants $V_{Na}$, $V_K$ and $V_L$ are the reversal potentials at which each current is balanced by the ionic concentration difference across the membrane. The activation variable $m_i$ of the sodium current $I_{Na,i}$ current is assumed to be fast and to always take its steady function
$m_{\infty}(v_i)$. On the other hand, both the inactivation variable $h_i$ of $I_{Na,i}$ and the activation variable $n_i$ of the potassium current $I_{K,i}$ obey a first-order kinetics of Eq.~(\ref{eq:CWBB}). Here, $\alpha_x$ and $\beta_x$ are the opening and closing rates of the gate variable $x (=h$ or $n$), respectively, and $\phi_x$ is the (dimensionless) temperature-like time scale factor.
Each FS Wang-Buzs\'{a}ki interneuron is also stimulated by using the common DC current $I_{DC}$ and an independent Gaussian white noise $\xi_i$ [see the 2nd and the 3rd terms in Eq.~(\ref{eq:CWBA})] satisfying $\langle \xi_i(t) \rangle =0$ and $\langle \xi_i(t)~\xi_j(t') \rangle = \delta_{ij}~\delta(t-t')$. We use following parameters for the FS Wang-Buzs\'{a}ki interneuron; $C=1 ~ \mu \textrm{F}/\textrm{cm}^2$, $g_{Na}=35 ~ \textrm{mS}/\textrm{cm}^2$, $g_{K}=9 ~ \textrm{mS}/\textrm{cm}^2$, $g_{L}=0.1 ~ \textrm{mS}/\textrm{cm}^2$, $V_{Na}=55 ~ \textrm{mV}$, $V_{K}=-90 ~ \textrm{mV}$, $V_{L}=-65 ~ \textrm{mV}$ and $\phi_h = \phi_n = 5$.
The last term in Eq.~(\ref{eq:CWBA}) represents the synaptic coupling of the network. Each neuron is connected to all the other ones through global synaptic couplings. $I_{syn,i}$ of Eq.~(\ref{eq:CWBF}) represents such synaptic current injected into the $i$th neuron. Here the coupling strength is controlled by the parameter $J$ and $V_{syn}$ is the synaptic reversal potential. We use $V_{syn} = -75$ mV for the inhibitory synapse. The synaptic gate variable $s$ obeys the first-order kinetics of Eq.~(\ref{eq:CWBC}) \citep{Order3,KI2}. Here, the normalized concentration of synaptic transmitters, activating the synapse, is assumed to be an instantaneous sigmoidal function of the membrane potential with a threshold $v^*$ in Eq.~(\ref{eq:CWBN}), where we set $v^* = 0$ mV and $\delta = 2$ mV. The transmitter release occurs only when the neuron emits a spike (i.e., its potential $v$ is larger than $v^*$). The synaptic channel opening rate, corresponding to the inverse of the synaptic rise time $\tau_r$, is $\alpha = 12 ~\textrm{ms}^{-1}$, which assures a fast rise of $I_{syn}$ \citep{KI2}. On the other hand, the synaptic closing rate $\beta$, which is the inverse of the synaptic decay time $\tau_d$, depends on the type of the synaptic receptors. For the inhibitory GABAergic synapse (involving the $\rm {GABA_A}$ receptors) we set $\beta = 0.1~\textrm{ms}^{-1}$ \citep{KI2}. Thus, $I_{syn}$ decays slowly.

\begin{figure}
\includegraphics[width=\columnwidth]{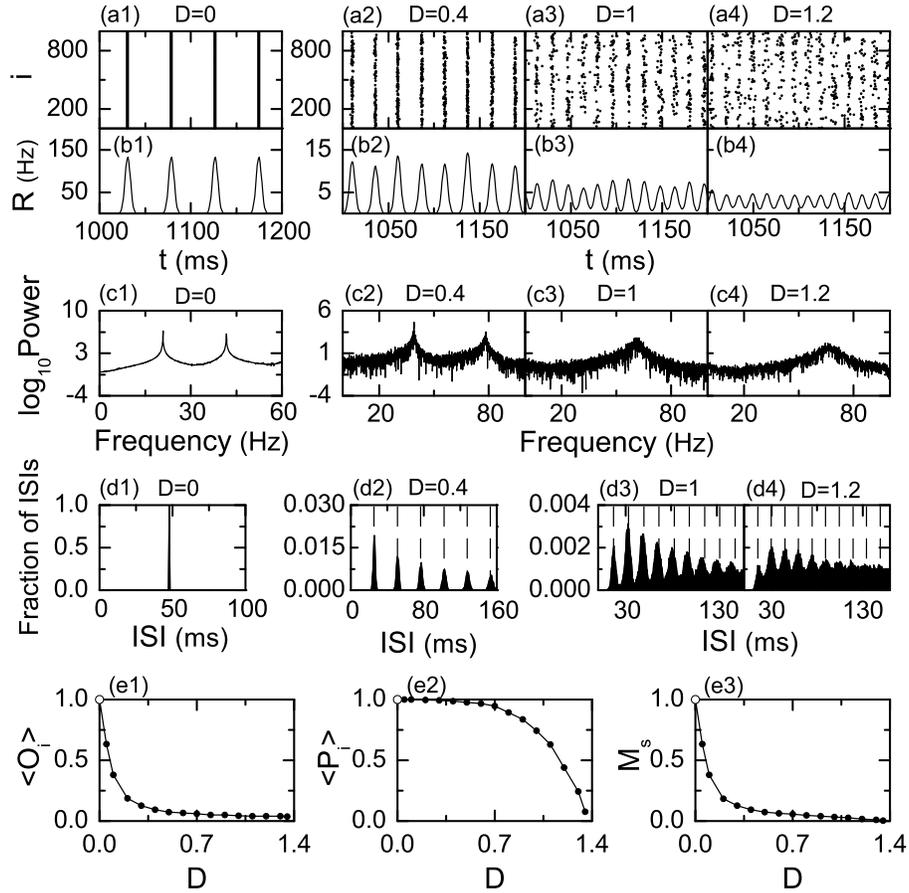}
\caption{Characterization of population spike synchronization in terms of the realistic statistical-mechanical spiking measure $M_s$, based on the IPSR kernel estimate $R(t)$, in an inhibitory population of $N$ $(=10^3)$ globally-coupled FS Wang-Buz\'{a}ki suprathreshold interneurons for $I_{DC}=2$ and $J=5$.  (a) Raster plots of neural spikes, (b) IPSR kernel estimates $R(t)$, (c) power spectra, and (d) ISI histograms for (a1)-(d1) $D=0$, (a2)-(d2) $D=0.4$, (a3)-(d3) $D=1$, and (a4)-(d4) $D=1.2$. In (c) each power spectrum is made of $2^{16}$ data points. In (d), each ISI histogram is made of $5 \times 10^4$ ISIs and the bin size is 3 ms. Vertical dashed lines in (d) represent the integer multiples of the global period $T_G$ of $R(t)$; $T_G$ = (d1) 47.6 ms, (d2) 25.5 ms, (d3) 16.7 ms, and (d4) 14.9 ms. (e1) Plot of $\langle O_i \rangle$ (average occupation degree of spikes) versus $D$. (e2) Plot of $\langle P_i \rangle$ (average pacing degree of spikes) versus $D$. (e3) Plot of $M_s$ (realistic ``statistical-mechanical'' spiking measure) versus $D$. To obtain $\langle O_i \rangle$, $\langle P_i \rangle$, and $M_s$ in (e1)-(e3), we follow the $3 \times 10^3$ stripes for each $D$. Open circles in (e1)-(e3) denote the data for $D=0$.}
\label{fig:WB_SM}
\end{figure}

Wang and Buzs\'{a}ki studied gamma rhythm by synaptic inhibition in their above hippocampal interneuronal network model \citep{KI2}. Here, we investigate  population neural synchronization for $I_{DC}=2$ (corresponding to a suprathreshold case where single Wang-Buzs\'{a}ki interneuron may fire spontaneously) and $J=5$ by varying the noise intensity $D$.  
Figure \ref{fig:WB_SM} shows the raster plots of spikes [(a1)-(a4)], the IPSR kernel estimates $R(t)$ [(b1)-(b4)], the power spectra of $R(t)$ [(c1)-(c4)], and the ISI histograms [(d1)-(d4)] for the synchronized cases of $D=0$, 0.4, 1, and 1.2 when $N=10^3$. We also measure the degree of population spike synchronization in terms of the average occupation degree $\langle O_i \rangle$, the average pacing degee $\langle P_i \rangle$, and the realistic statistical-mechanical IPSR-based spiking measure $M_s$ for 16 values in the synchronized regime, and the results are shown in Figs.~\ref{fig:WB_SM}(e1)-\ref{fig:WB_SM}(e3). 
For $D=0$, suprathreshold Wang-Buzs\'{a}ki interneurons exhibit complete synchronization with $\langle O_i \rangle=1$, $\langle P_i \rangle=1$, and $M_s = 1$ [see open circles in Figs.~\ref{fig:WB_SM}(e1)-\ref{fig:WB_SM}(e3)], in contrast to the case of subthreshold Izhikevich interneurons. Clear fully-occupied stripes appear in the raster plot of spikes, as shown in Fig.~\ref{fig:WB_SM}(a1). Then, the IPSR kernel estimate $R(t)$ exhibits regular population oscillation of 21 Hz,  as shown in Figs.~\ref{fig:WB_SM}(b1) and \ref{fig:WB_SM}(c1). Obviously, the ISI histogram has a single peak at $T_G$ (=47.6 ms) [see Fig.~\ref{fig:WB_SM}(d1)]. Hence, all FS Wang-Buzs\'{a}ki suprathreshold interneurons keep perfect pace with other ones. 
However, as $D$ begins to increase, an abrupt decrease in the average occupation degree $\langle O_i \rangle$ occurs, while the average pacing degree $\langle P_i \rangle$ is still nearly the same [see Figs.~\ref{fig:WB_SM}(e1)-\ref{fig:WB_SM}(e2)]. As an example, we consider the case of $D=0.4$ with $\langle O_i \rangle = 0.094$, $\langle P_i \rangle = 0.99$, and $M_s=0.093$. As shown in Fig.~\ref{fig:WB_SM}(a2), the density of stripes (denoting the occupation degree) is much decreased, while only a little smearing of the stripes (representing the pacing degree) occurs. As in the case of the Izhikevich interneuron model, the partial occupation occurs due to stochastic spike skipping of individual interneurons seen well in the ISI histogram of Fig.~\ref{fig:WB_SM}(d2) with clear (well-separated) multiple peaks appearing at multiples of the period of $T_G$ (=25.5 ms) of the IPSR kernel estimate $R(t)$. Then, the IPSR kernel estimate $R(t)$ shows relatively regular oscillation with much decreased amplitude but faster population frequency of 39.2 Hz [see Fig.~\ref{fig:WB_SM}(c2)]. As $D$ is further increased and becomes larger than about 0.7, a rapid  decease in the average pacing degree $\langle P_i \rangle$ also begins to occur [see Fig.\ref{fig:WB_SM}(e2)]. For example, stripes in the raster plots of Figs.~\ref{fig:WB_SM}(a3)-\ref{fig:WB_SM}(a4) for $D=1$ and 1.2 become more and more smeared, and hence the average pacing degree $\langle P_i \rangle$ is decreased rapidly  with increasing $D$. This smearing of stripes can be understood from the change in the structure of the ISI histograms, as shown in Figs.~\ref{fig:WB_SM}(d3)-\ref{fig:WB_SM}(d4). Merging of peaks results in the smearing of stripes, leading to rapid decrease in $\langle P_i \rangle$. Eventually, when passing the higher threshold $D^*$ ($\simeq 1.38$), unsynchronized states appear due to complete smearing of stripes (i.e., stripes no longer appear in the raster plot of spikes).

\begin{figure}
\includegraphics[width=\columnwidth]{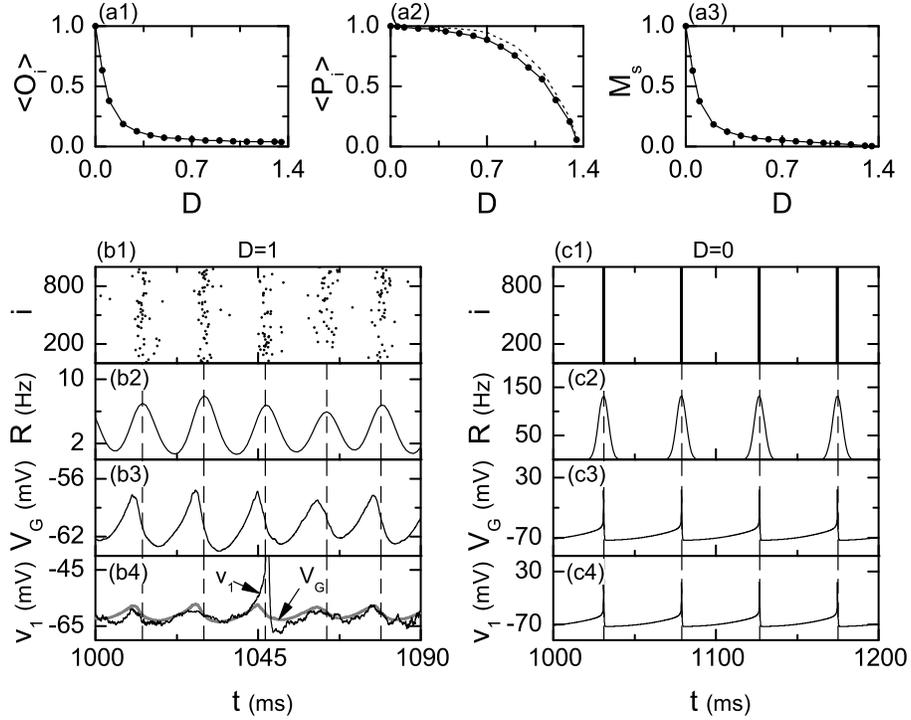}
\caption{Characterization of population spike synchronization in terms of the conventional statistical-mechanical $V_G$-based measures in an inhibitory population of $N$ $(=10^3)$ globally-coupled FS Wang-Buzs\'{a}ki suprathreshold interneurons for $I_{DC}=2$ and $J=5$. (a1) Plot of $\langle O_i \rangle$ (average occupation degree of spikes) versus $D$. (a2) Plot of $\langle P_i \rangle$ (average pacing degree of spikes) versus $D$ [for comparison, $\langle P_i \rangle$ in Fig.~\ref{fig:WB_SM}(d2), based on IPSR, is shown with the dotted line]. (a3) Plot of $M_s$ (conventional ``statistical-mechanical'' spiking measure) versus $D$. To obtain $\langle O_i \rangle$, $\langle P_i \rangle$, and $M_s$ in (a1)-(a3), we follow the $3 \times 10^3$ stripes for each $D$. (b1) Raster plot of neural spikes, (b2) IPSR kernel estimate $R(t)$, (b3) global potential $V_G$, and (b4) membrane potential $v_1$ of the first neuron for the weak sparse synchronization when $D=1$. (c1) Raster plot of neural spikes, (c2) IPSR kernel estimate $R(t)$, (c3) global potential $V_G$, and (c4) membrane potential $v_1$ of the first neuron for the strong spike synchronization when $D=0$. Vertical dashed lines in (b2)-(b4) and (c2)-(c4) represent maxima of $R(t)$.}
\label{fig:SM_VG}
\end{figure}

Finally, to examine the accuracy of the realistic statistical-mechanical IPSR-based measures, we  characterize neural synchronization in the above Wang-Buzs\'{a}ki interneuron model in terms of the conventional statistical-mechanical $V_G$-based measures. The results of $\langle O_i \rangle$, $\langle P_i \rangle$, and $M_s$ are given in Figs.~\ref{fig:SM_VG}(a1)-\ref{fig:SM_VG}(a3), respectivley. The values of average occupation degree $\langle O_i \rangle$ are nearly the same for both cases of the realistic and conventional measures. However, some discrepancies occur for the values of average pacing degree $\langle P_i \rangle$ [compare the solid and dotted lines in Fig.~\ref{fig:SM_VG}(a2)]. For the case of weak sparse spike synchronization, the values of $\langle P_i \rangle$ for the case of the conventional $V_G$-based measures are a little smaller than those for the case of the realistic IPSR-based measures, although the values of $\langle P_i \rangle$ for both kinds of measures are nearly the same for the case of strong spike synchronization sufficiently near $D=0$.
As an example of weak sparse synchronization, we consider the case of $D=1$.  We note that IPSR reflects spike synchronization in the raster plot of spikes, while the ensemble-averaged $V_G$ shows synchronization of individual membrane potentials. As shown in Fig. \ref{fig:SM_VG}(b1)-\ref{fig:SM_VG}(b3), maxima of IPSR appear at the times with highest spike density in the raster plot of spikes, but maxima of $V_G$ appear a little earlier than peaks of IPSR. When the occupation and pacing degrees are low, contributions of individual spikings to $V_G$ is negligibly small, and hence $V_G$ is formed mainly by contributions of subthreshold oscillations [see Fig. \ref{fig:SM_VG}(b4)]. Thus, peaks of $V_G$ appear nearly at maxima of small subthreshold oscillations of individual membrane potentials. In each individual potential, spikings occur when subthreshold oscillation is more increased and passes a threshold, and hence they appear a little behind maxima of subthreshold oscillations. Thus, maxima of $V_G$ (mainly contributed by small subthreshold oscillations) appear a little earlier than peaks of IPSR. Hence, in the case of weak sparse synchronization, contributions of spikes in the raster plots to $V_G$ is a little lower than those to IPSR, and the values of $\langle P_i \rangle$ for the conventional $V_G$-based case become smaller than those for the IPSR-based case, unlike the case of strong spike synchronization. Consequently, more accurate characterization of weak sparse spike synchronization can be achieved in terms of the realistic IPSR-based measure, in comparison with the conventional $V_G$-based measure. 
As an example of strong synchronization, we consider the case of $D=0$. When both the occupation and pacing degree are high, maxima of $V_G$ coincide well with those of IPSR because peaks of $V_G$ are formed mainly by contributions of individual spikings [see Figs.~\ref{fig:SM_VG}(c1)-\ref{fig:SM_VG}(c4)]. Thus, both the realistic IPSR-based and conventional $V_G$-based measures give nearly the same accurate characterization of strong spike synchronization, unlike the case of weak sparse synchronization.

\section{Summary} \label{sec:SUM}
The experimentally-obtainable IPSR is a realistic population quantity which is appropriate for describing collective behavior in both experimental and computational neuroscience. Instead of the ensemble-averaged potential $V_G$ which is often used in computational and theoretical neuroscience, we use the IPSR kernel estimate $R(t)$ and develop realistic thermodynamic and statistical-mechanical measures, based on $R(t)$, for characterization of neural synchronization in an inhibitory population of FS Izhikevch subthreshold interneurons. 
The range of noise intensity where synchronized neural oscillations occur has been determined through calculation of the realistic thermodynamic order parameter $\tilde{\cal{O}}$. 
For the synchronous case, we have characterized the degree of population spike synchronization seen in the raster plot of spikes in terms of realistic statistical-mechanical spiking measure $M_s$ by taking into consideration both the occupation and the pacing degrees of spikes in the raster plot. 
Particularly, the pacing degree between spikes is determined in a statistical-mechanical way by quantifying the average contribution of (microscopic) individual spikes to the (macroscopic) IPSR kernel estimate $R(t)$. 
In a statistical-mechanical sense, our realistic spiking measure $M_s$ supplements the conventional microscopic spiking measures, based on the spike- and the ISI-distances \citep{SD1,SD2,SD3,ISID,SD4,SD5}.
Thus, we have explicitly shown the usefulness of the realistic measures $\tilde{\cal{O}}$ and $M_s$ for characterization of sparsely synchronized cortical rhythms which have partially-occupied stripes in the raster plot. 
For examination on usefulness of realistic statistical-mechanical measure $M_s$, we have also successfully characterized neural synchronization in another population of FS Wang-Buzs\'{a}ki suprathrshold interneurons. 
Furthermore, it has been shown that, in comparison with the conventional statistical-mechanical $V_G$-based measures, more accurate characterization of weak sparse spike synchronization can be achieved in terms of statistical-mechanical IPSR-based measures. 
We expect that $M_s$ might be easily implemented to quantify not only the degree of population spike synchronization in an experimentally-obtained raster plot of neural spikes, but also the reliability of spike timing and stimulus discrimination in real experimental data, as discussed below. 

\begin{figure}
\includegraphics[width=\columnwidth]{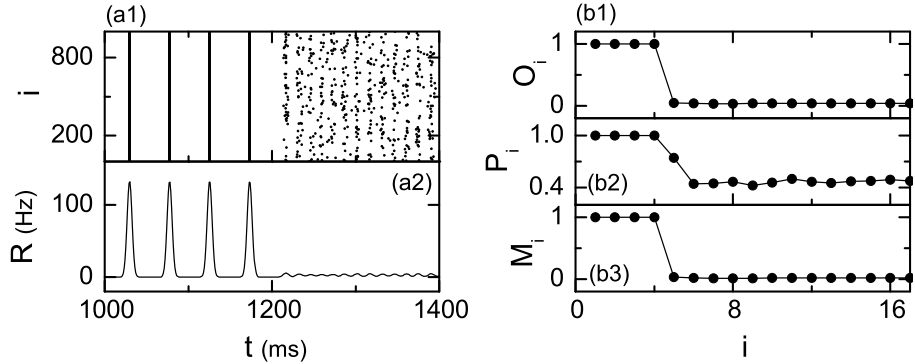}
\caption{Characterization of the dynamics of synchronization in terms of the realistic statistical-mechanical IPSR-based measures in an inhibitory population of $N$ $(=10^3)$ globally-coupled FS Wang-Buzs\'{a}ki suprathreshold interneurons for $I_{DC}=2$ and $J=5$. At $t=1200$, the value of $D$ is increased from $D=0$ to $D=1.2$. (a1) Raster plots of neural spikes and (a2) IPSR kernel estimate $R(t)$. Plots of (b1) $O_i$ (occupation degree of spikes in the $i$th stripe),  (b2) $P_i$ (pacing degree of spikes in the $i$th stripe), and (b3) $M_i$ (spiking measure in the $i$th stipe) versus $i$ (stripe).}
\label{fig:SM_Dyn}
\end{figure}

Application of realistic statistical-mechanical measures to real experimental data is beyond the scope of present work, and hence we leave such applications as future works. 
Here, we just discuss some ``possibilities'' of these applications to real experimental data. 
For characterization of neural synchronization in real experiments, one might get a raster plot of neural spikes through multi-unit recordings and spike sorting \citep{Exp}, obtain smooth IPSR kernel estimate $R(t)$, and then characterize population spike synchronization in terms of realistic statistical-mechanical measure, based on $R(t)$. As a second example, we consider spike-timing reliability \citep{ST1,ST2,ST3,ST4,ST5,ST6,ST7,ST8,ST9}. One might obtain spike trains via single-unit recordings in response to repeated trials of presenting the same stimulus to a single neuron, forms a raster plot of spike trains (trials versus spike times) and the smooth kernel estimate of instantaneous firing rate $R(t)$ [corresponding to smooth peri-stimulus time histogram (PSTH)], and then apply realistic statistical-mechanical measure for characterization of spike-timing reliability.
For this case, average occupation degree $\langle O_i \rangle$ and pacing degree $\langle P_i \rangle$ correspond to conventional spike-timing reliability and precision, respectively. We also discuss the stimulus discrimination \citep{STD1,STD2,Song1,Song2,STD3,SD}. As an example, we consider 20 songs for classification in songbirds \citep{Song1,Song2}. For each song, one might obtain spike trains via single-unit recordings in response to 10 repeated trials of presenting the same song and forms a ``template" raster plot of spike trains and the corresponding smooth kernel estimate $R(t)$ of instantaneous firing rate for the song. Thus, we have 200  spike trains and obtain 20 template kernel estimates $R(t)$. Our template $R(t)$ for each song is a ``macroscopic'' one, in contrast to the conventional randomly-chosen ``microscopic" template spike-train \citep{Song1,Song2}. 
Then, we obtain statistical-mechanical pacing degrees between each spike train and 20 template kernel estimates $R(t)$, assign each spike train to the closest template, and compute correct percentage. 
In this way, we believe that the statistical-mechanical pacing degree could be used as a ``similarity'' measure to quantify neural discrimination.
Finally, we emphasize that realistic statistical-mechanical measures ($O_i$, $P_i$, and $M_i$), based on $R(t)$, are computed for each $i$th global cycle of $R(t)$ (corresponding to $i$th stripe in the raster plot of spikes), as shown in Fig.~\ref{fig:SM1}. 
Hence, changes in the level of synchronization (i.e., dynamics of synchronization) can also be investigated in terms of realistic statistical-mechanical measures. 
In both the Izhikevich and the Wang-Buzs\'{a}ki neuron model, the external stimulus is given by $I_{ext} = I_{DC} + D \xi$ [see Eqs.~(\ref{eq:CIZA}) and (\ref{eq:CWBA})]. Here, the constant DC current $I_{DC}$ corresponds to average value of $I_{ext}$, and the fluctuation degree of $I_{ext}$ is controlled by the noise intensity $D$. 
We consider the following hypothetical situation in the Wang-Buzs\'{a}ki interneuron model. At a specific time $(t=1200)$, the value of $D$ is assumed to increase from $D=0$ to $D=1.2$ as a result of some ``event." 
Figure \ref{fig:SM_Dyn} shows raster plot of spikes, $R(t)$, $O_i$, $P_i$, and $M_i$. 
For $t<1200$, clear fully-occupied stripes exist in the raster plot of spikes and the corresponding IPSR $R(t)$ exhibits regular large-amplitude oscillation with population frequency $f_p=21$ Hz, as shown in Figs.~\ref{fig:SM_Dyn}(a1)-\ref{fig:SM_Dyn}(a2). 
However, for $t>1200$, smeared partially-occupied stripes appear [see Fig.~\ref{fig:SM_Dyn}(a1)], and the corresponding $R(t)$ shows very small-amplitude oscillation with increased frequency $f_p=67$ Hz, as shown in Fig.~\ref{fig:SM_Dyn}(a2).
We characterize the dynamics of synchronization in terms of the realistic statistical-mechanical IPSR-based spiking measure.
For $t<1200$, complete synchronization occurs with $O_i=1$, $P_i=1$, and $M_i=1$. 
However, for $t>1200$, remarkable decrease in $O_i$ ($\langle O_i \rangle = 0.039$), $P_i$ ($\langle P_i \rangle = 0.44$), and $M_i$ ($\langle M_i \rangle = 0.017$) occurs. 
Based on these results, we believe that time courses of kernel estimate $R(t)$ and realistic statistical-mechanical measures might be usefully used for characterization of change in both the population frequency of synchronized rhythm and the synchronization degree occurring in event-related synchronization/desynchronization \citep{ERS1,ERS2,ERS3}. 

\section*{Acknowledgments}
W. Lim acknowledges financial support from the Daegu National University of Education (Grant No. RC2013063).

\end{document}